\documentclass{article}
     \PassOptionsToPackage{numbers, compress}{natbib}

     \usepackage[preprint]{neurips_2022}

\usepackage[utf8]{inputenc} 
\usepackage[T1]{fontenc}    
\usepackage{hyperref}       
\usepackage{url}            
\usepackage{booktabs}       
\usepackage{amsfonts}       
\usepackage{nicefrac}       
\usepackage{microtype}      
\usepackage{xcolor}         
\usepackage{pdfpages}
\title{Time Series Synthesis via Multi-scale Patch-based Generation of Wavelet Scalogram}

\author{%
  Amir Kazemi \\
  Department of Civil and Environmental Engineering\\
  University of Illinois at Urbana-Champaign\\
  Urbana, IL 61801 \\
  \texttt{kazemi2@illinois.edu} \\
  \And
  Hadi Meidani \\
  Department of Civil and Environmental Engineering\\
  University of Illinois at Urbana-Champaign\\
  Urbana, IL 61801 \\
  \texttt{meidani@illinois.edu} \\
}

\begin{document}

\maketitle

\begin{abstract}
A framework is proposed for the unconditional generation of synthetic time series based on learning from a single sample in low-data regime case.
The framework aims at capturing the distribution of patches in wavelet scalogram of time series using single image generative models and producing realistic wavelet coefficients for the generation of synthetic time series.
It is demonstrated that the framework is effective with respect to fidelity and diversity for time series with insignificant to no trends.
Also, the performance is more promising for generating samples with the same duration (reshuffling) rather than longer ones (retargeting).

\end{abstract}

\section{Introduction}

A time series is a sequence of observations in chronological order which typically represents the temporal alteration of a physical or abstract quantity. The available observation of time series are scarce in many applications such as healthcare \cite{esteban2017real}, finance \cite{assefa2020generating}, and industries such as smart grids \cite{zhang2018generative}, power and energy systems \cite{li2020creation,talbot2020correlated}, and networks and service management \cite{wang2021tsagen}. This could be due to technical limitation in instrumentation and measurement, or  due to privacy, preventing real data to be available for model training \cite{el2020practical}.  This is while typically many samples of time series are needed for an effective approximation of the underlying random process. Therefore in the absence of sufficient observation data, to  enhance  downstream tasks like classification and clustering, an effective solution is  generating synthetic time series.

Generative Adversarial Networks (GAN), as one of the  most popular generative frameworks in the recent years, have been used and tailored for generating synthetic time series \cite{brophygenerative}. While GAN-based time-series generation like C-RNN-GAN\cite{mogren2016c}, RCGAN\cite{esteban2017real}, and TimeGAN \cite{yoon2019time} are not substantially less data-hungry than GAN, the advent of single image generation architectures inspires the idea of capturing the recurrence of data patterns within individual time series. This work is an effort to employ existing single image generation architectures for the generation of synthetic time series.  Therefore, a brief background on single image generative models is covered before elaborating on this contribution.

GAN adopts a game-theoretic deep learning approach to capture data distribution in an implicit way. The adversarial framework consists of a pair of networks called generator and discriminator. The generator produces fake data from a latent space with an aim to deceive the discriminator in believing the generated samples are real. The discriminator returns a score which shows the probability of being real for each sample \cite{goodfellow2014generative}. A problem with GANs is that when the distance between the distributions of real and fake data is too large, the gradient of the loss function produces random directions \cite{karras2018progressive}. While the choice of distributional distance may alleviate this, an alternative approach to improve the quality, variability, and the stability of the framework is to train the networks in a multi-scale manner. To this end, a multi-scale version of GAN, called Progressive growing of GANs (PGGAN) has been developed based on  the idea of growing generator and discriminator \cite{karras2018progressive}. PGGAN initiates  a low-resolution generator and discriminator which collectively  unveil the global structure of the image. As training proceeds, the networks are grown gradually and simultaneously to discover more details from high-resolution images. The idea of such a pyramidal hierarchy has been applied to generating images from a single one by exploiting the recurrence of internal information of only one image in an architecture called SinGAN \cite{shaham2019singan}. The same pyramidal hierarchy was later applied to non-adversarial architectures \cite{granot2022drop,elnekave2022generating}.

\paragraph{Contribution.} As mentioned earlier, it is intuitive to look at a sample of time series as a one-shot target, just as the single image generation architectures synthesize the texture of an individual image. 
In order to produce synthetic time series in the same fashion, we seek to work with the transformation of time series in the time-frequency domain (spectrogram) or the time-scale domain (wavelet scalogram), and exploit this information in single image generative models. The mentioned transforms unveil characteristics of a time series in two domains. In particular, a scalogram  captures the features of real-world time series at various scales of mother wavelet and times. Therefore, we employ wavelet transform here as a preprocessing step to obtain the scalogram and its inverse as a post-processing step. We apply a single-image generator on the wavelet scalograms of individual time series to generate patch-similar scalograms. Generation is performed for targets of the same temporal duration, i.e. reshuffling, and longer duration, called retargeting. We then obtain performance metrics of fidelity and diversity for the generated results using three sets of random processes as a proof of concept for the idea. One-shot generation of time series leverages the repeated patterns of individual samples for low-data regime applications.

\section{Framework}
\begin{figure}
  \centering
  \includegraphics[width=3in]{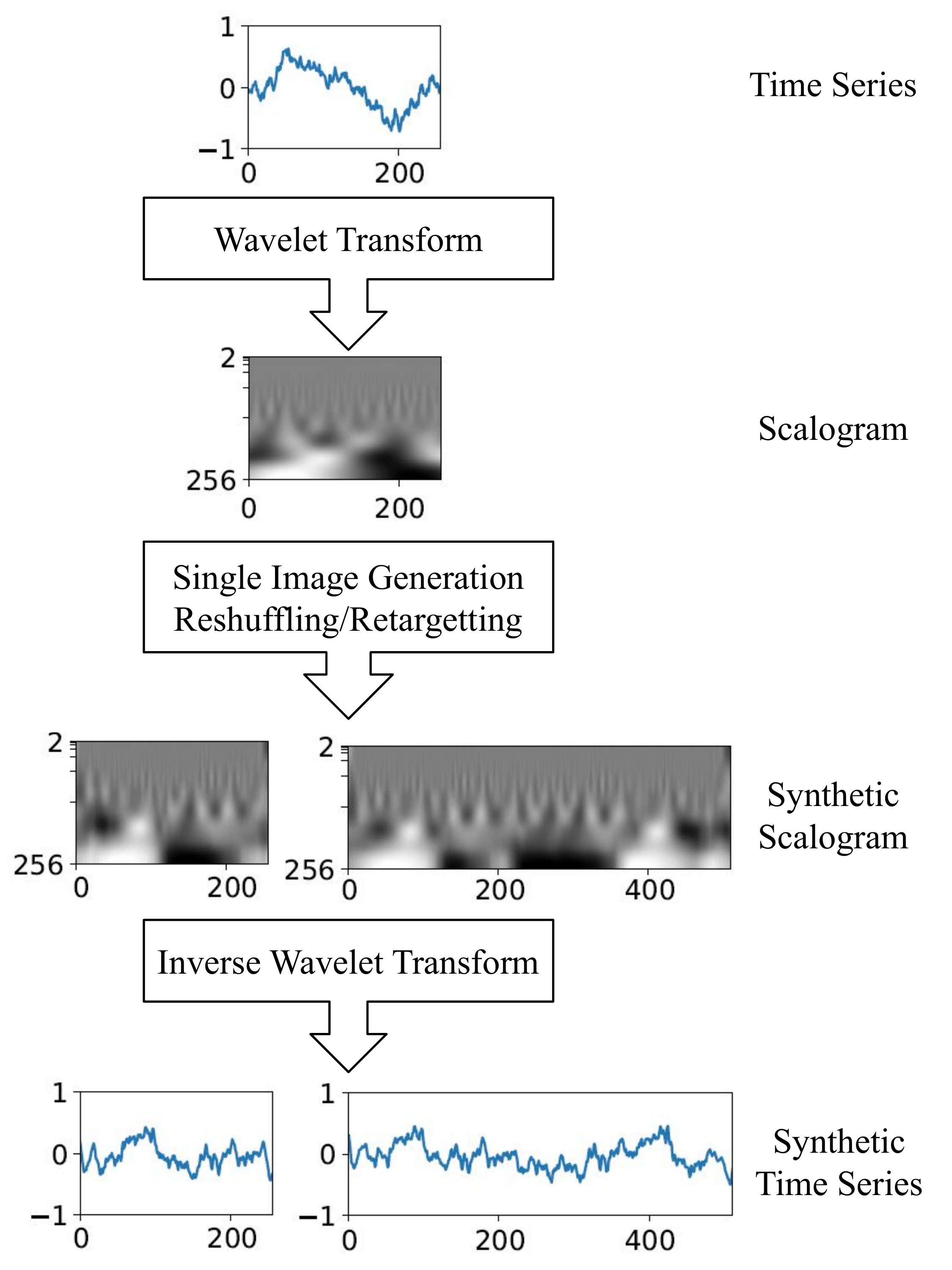}
  \caption{Multi-scale patch-based generation of wavelet scalogram for synthetic time series}
  \label{framework}
\end{figure}

The framework for the generation of synthetic time series from a single time series consists of the following three steps as illustrated in Figure \ref{framework}.

\paragraph{Wavelet transform.}
A wavelet transform is applied on the input (single)  time series. Because time series in most of the applications are  real-valued, the coefficients of wavelet transform are real-valued if a real mother wavelet is applied. This implies that wavelet coefficients can be represented by a single channel 2D tensor or image (a complex wavelet transform otherwise produces two channels for real and complex parts of the wavelet coefficient). Note that the scalogram  traditionally consists of the absolute value of the wavelet coefficient, but in the context of this work the scalograms contain  normalized values of the signed coefficients. Wavelet transform, not only provides a 2D image representation of the time series, but also reveals how the frequency content of the series varies along time.

\paragraph{Generating wavelet scalograms.}
In order to generate synthetic wavelet scalograms, we apply a single image generative model to resuhuffle or retarget the obtained wavelet scalogram along time dimension. These generative models, as reviewed in the introduction, have a pyramidical hierarchy of generators which learns the repeating patterns of the image in different scales or resolutions. Note that the scale in the term "multi-scale patch-based generation" refers to the resolution of the scalogram images and should be distinguished from the scale of mother wavelets.

We may choose any of the single image generative architectures in the literature. As a pioneering single image generative architecture, SinGAN  \cite{shaham2019singan} adopts the idea of GAN by considering multiple patches within a single image, instead of training a network on multiple images. Recently, a novel generative multiscale algorithm, namely Generative Patch Nearest-Neighbor (GPNN) \cite{granot2022drop} argued that  GAN-based single image generators  (1) have long computational time for training, (2) suffer from optimization traps such as mode collapse, and (3) produce relatively poor visual quality results compared to  non-deep learning multiscale methods. While these drawbacks for single-image learning are attributed to GANs, the unconditional feature of GAN is exploited in GPNN for the sake of generation diversity. Therefore, the coarse-scale phase of GPNN employs a noise feed to help the diversity of generated samples, while the optimization-free nature of patch similarity measure makes GPNNs substantially faster and more stable compared to other methods such as SinGAN. The patch-similarity measure in GPNN has recently been challenged and substituted by a faster method namely direct patch distributions matching (GPDM) which uses sliced Wasserstein distance (SWD) instead of bidirectional similarity (BDS) \cite{elnekave2022generating}. Based on these features, in this work, GPDM is employed for the generation of synthetic wavelet scalograms.

\paragraph{Inverse wavelet transform.}
The generated scalograms are transformed to time domain using the inverse wavelet transform. This lead to the generation of synthetic time series whose wavelet scalograms are similar, with respect to patches in different resolutions, to the scalogram of the original time series.

\section{Experiments}

\paragraph{Time Series Models.}
As the underlying process, three random processes are considered: the Brownian bridge, the drifted Brownian motion, and the Wiener process. Corresponding to each random process, datasets of different sizes are generated for time series of length 256. The objective is to obtain synthetic datasets of size $5000\times256$ using datasets of size $n\times256$ with $n\in\{5,50,500\}$ and to compare the results with the ground truth datasets of  size $5000\times256$. Time series are generated using \url{https://github.com/crflynn/stochastic}. Also, to demonstrate the capability of the proposed framework in generating longer time series, we used the model to generate time series of length 512 using datasets of size $n\times256$ and compared them to a ground truth dataset of size $5000\times512$. Note that to reach $5000$ synthetic time series using $n$ original samples, $5000/n$ samples is generated for each of them.

\paragraph{Wavelet Transform.}
For the wavelet transform, the mother wavelet is the real Morlet and the scales are dyadic meaning that the scales follow a geometric sequence of ratio two, i.e. $[2^1,...,2^{8}]$. The wavelet transform is performed using \url{https://github.com/PyWavelets/pywt}. Note that for the sake of visual clarity, scalograms are resized with a different aspect ratio along scale size and time so that the variation of coefficients becomes more observable.

\paragraph{Generative Model.}
As mentioned in the introduction, adversarial and non-adversarial generative models have been proposed for single image generation in  recent years.
In this work, a non-adversarial fast model namely GPDM is employed to generate patch-similar wavelet scalograms of time series; see \url{https://github.com/ariel415el/GPDM} \cite{elnekave2022generating}.

\paragraph{Metrics of Performance.}
The fidelity and diversity of generated samples from generative models can be evaluated by different metrics. Although the Frechet Inception Distance (FID) and Inception Score (IS) have been frequently  used for such performance evaluation, the fact that they fail to distinguish fidelity and diversity aspects of the generation gave rise to other metrics like precision and recall. Therefore, the improved precision and recall are employed in this work to evaluate the fidelity and diversity of the generated time series, respectively \cite{sajjadi2018assessing}. The original implementation of the improved precision and recall can be found at \url{https://github.com/kynkaat/improved-precision-and-recall-metric} \cite{kynkaanniemi2019improved}, but the implementation at \url{https://github.com/clovaai/generative-evaluation-prdc} is preferred for time series  as it works with tensors and not necessarily images \cite{naeem2020reliable}. 

\begin{figure}
  \centering
  \includegraphics[width=11cm,trim={3cm 3cm 3cm 5cm}]{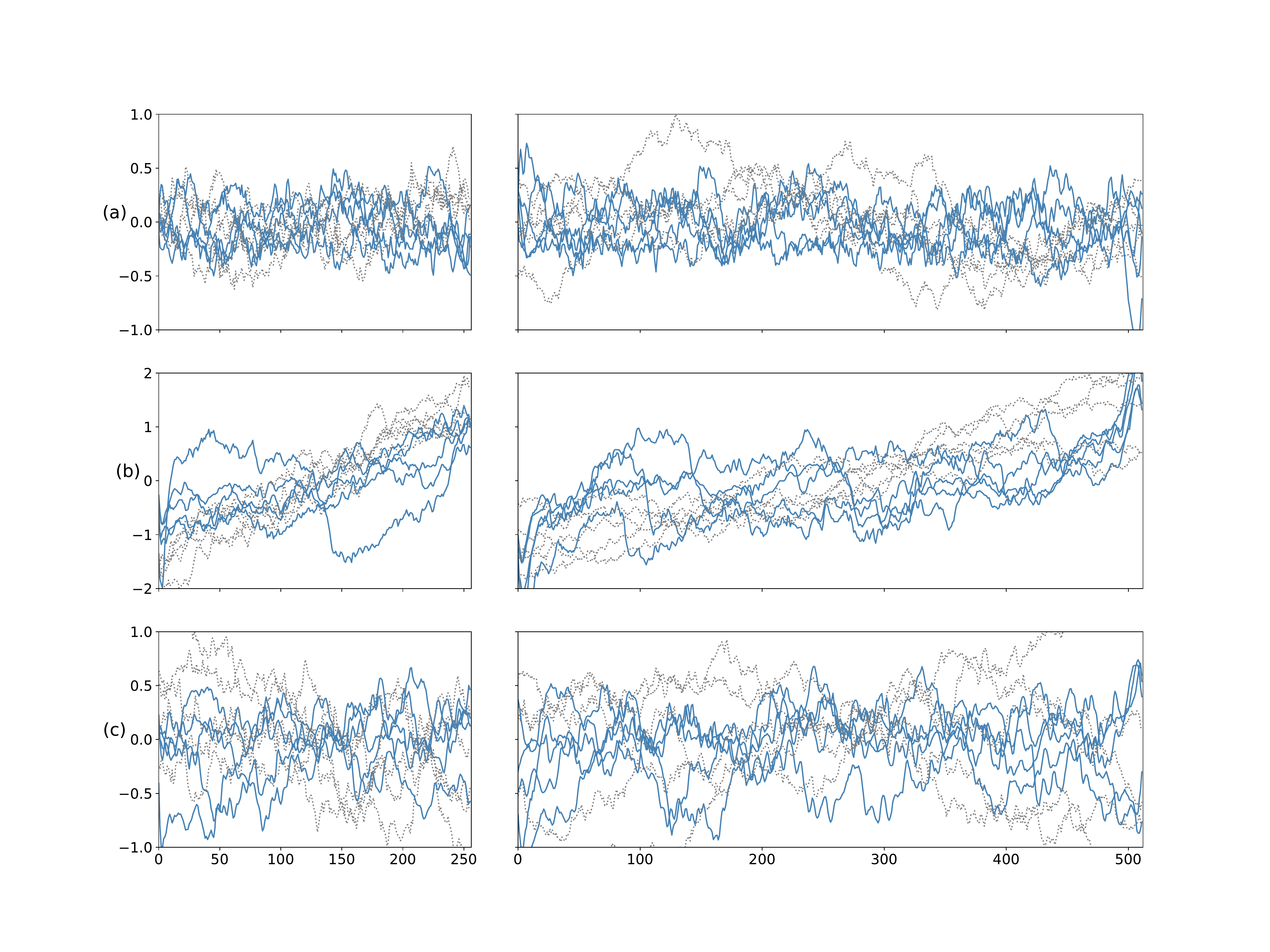}
  \caption{Illustration of synthetic time series (solid blue lines) versus real samples (gray dots) for (a) Brownian bridge, (b) drifted Brownian motion, and (c) Wiener process. Left and right columns show reshuffling and retargeting results against ground truth, respectively.}
  \label{results2}
\end{figure}

\paragraph{Results and Discussion.}
As mentioned earlier, three random processes are selected to generate samples.  Figure \ref{results2} shows samples of synthetic time series versus real ones. Each of the three processes has certain characteristics which reveal the capabilities and weak points of the proposed generative framework in its current form.  

\begin{figure}
  \centering
  \includegraphics[width=\textwidth,trim={4cm 4cm 4cm 4cm}]{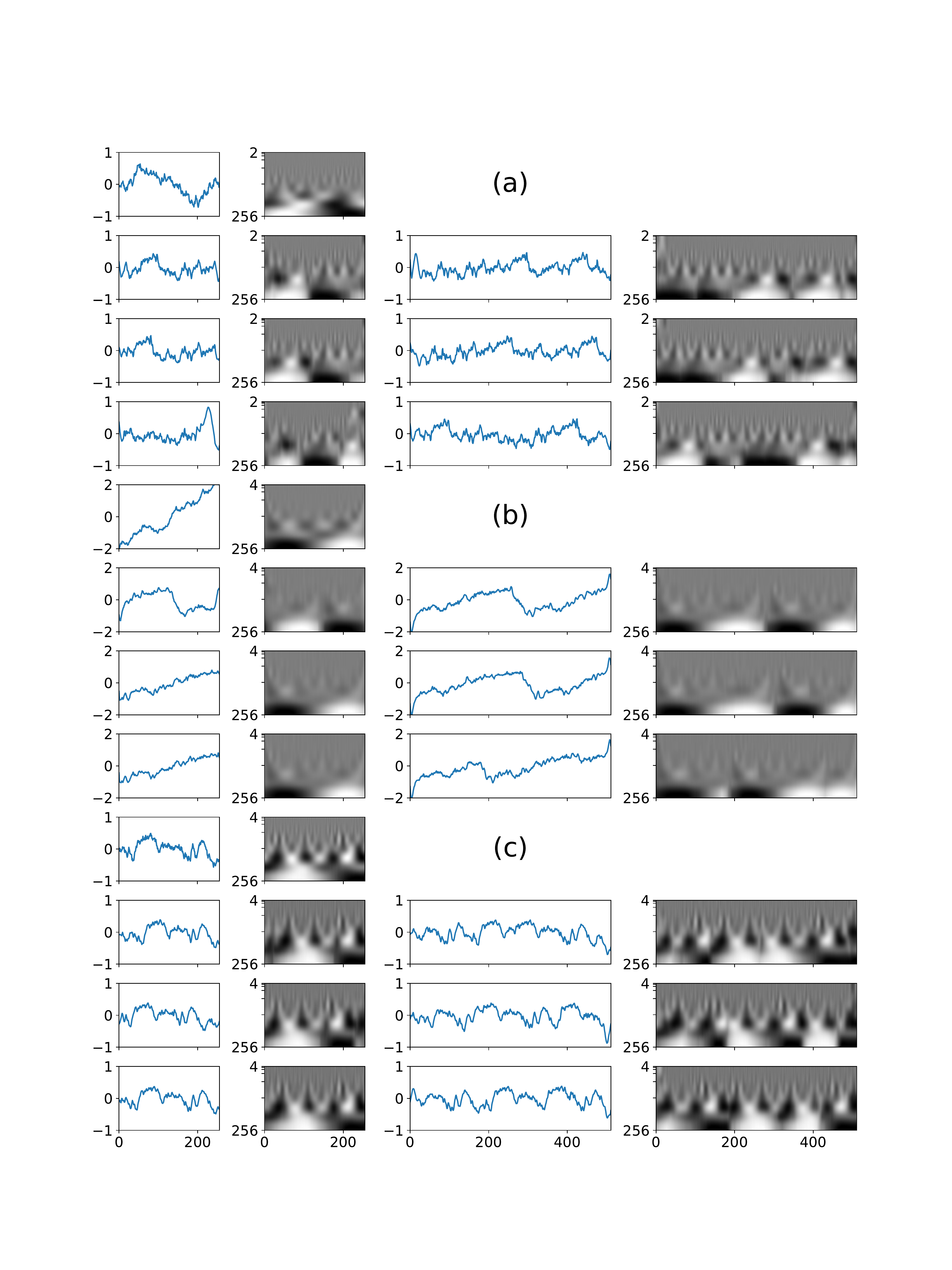}
  \caption{Time series and corresponding wavelet scalograms for three random process: (a) Brownian bridge, (b) drifted Brownian motion, and (c) Wiener process. The first row in each case is the real sample of the time series and the following three rows shows three samples of reshuffled and retargeted wavelet scalogram generating synthetic time series. Wavelet scalogram values are normalized to $[0,1]$ for the sake of visual clarity.}
  \label{results}
\end{figure}

The Brownian bridge is a Wiener process $W_t$ which is conditioned by $W_0=W_T$ implying the most uncertainty in the middle of $[0,T]$.
In the multi-scale patch-similar generation of wavelet scalogram for this process, the condition $W_0=W_T$ is not enforced explicitly.
Therefore it is expected that $W_0=W_T$ does not necessarily hold for the generated time series; yet Figure \ref{results}(a) demonstrates that the particular form of bridge is evident to some extent both for reshuffling and retargeting cases. Despite the success in capturing the bridge form, the inaccurate estimations of $W_t$ at $t=\{0,T\}$ and nearby points has lead to a lack of precision. Also for the retargeting case, the longer duration of the Brownian bridge should lead to more uncertainty in the middle of the bridge which is not realized here. This justifies the resulting low precision in Table \ref{retargeting}, compared to that in the reshuffling case in Table \ref{reshuffling}.

The Drifted Brownian motion is more challenging than the Brownian bridge as it is a typical trended process which produces wavelet scalograms with a minimal repeating pattern. While such a pattern is essential for a precise texture synthesis, the drifted Brownian motion trend as shown in Figure \ref{results}(b) lacks such a repeated texture in the wavelet scalogram.
Therefore, while the generated samples are locally very similar to the original sample, they lack a global similarity which ruins the trend.
Although hyper-parameter tuning of GPDM may lead to a more similar global structure of trended time series, this would be at the cost of less diversity in the samples. In other words, the well-known trade-off between precision and recall is more challenging for trended time series without substantial repeating pattern in the wavelet scalogram. This issue underscores the importance of detrending time series before using texture synthesis for generating their scalograms. By retargeting, instead of reshuffling, one may obtain even less precision as there is more room for unnecessary shuffling of the patches and loss of trends; see Tables \ref{reshuffling} and \ref{retargeting}.

Among the three models, the Wiener process is the least trended time series and therefore its generation precision is the highest in both reshuffling and retargeting. However, retargeting still suffers from lack of precision which is attributed to the influence of duration on the global structure of the time series; see Tables \ref{reshuffling} and \ref{retargeting}.
Such an adverse influence is more evident in strongly trended time series like the drifted Brownian motion.

The diversity of the generated time series are affected by two factors: the level of high-frequency content and the retargeting scale factor.
More high-frequency content implies more repeated pattern which leads to more generation diversity. As observed in Figure \ref{results}, the simulated Brownian bridge has the most high-frequency content, followed by the Wiener process and the drifted Brownian motion. This is supported by the the recall values for reshuffling and retargeting cases (see Tables \ref{reshuffling} and \ref{retargeting}). Also, retargeting gives more diverse samples compared with the reshuffling for the same process. That is because retargeting with a scale factor larger than one provides more temporal capacity for patch relocation.

It is also evident that increasing the size of training dataset provides more information to the generator to escape the mode collapse and therefore mostly results in higher recalls. Training size does not meaningfully affect precision as the generation fidelity is based on capturing the patterns within individual samples (Tables \ref{reshuffling} and \ref{retargeting}). Finally, Figure \ref{results2} shows  the fidelity and diversity of generated and real samples of time series.

\begin{table}
  \caption{Performance metrics for time series using reshuffling wavelet scalograms}
  \label{reshuffling}
  \centering
  \begin{tabular}{ccccccc}
    \toprule
    \multicolumn{3}{r}{Precision , Recall}                   \\
    \cmidrule(r){2-4}
    Size of Training Dataset & Brownian Bridge &  Drifted Brownian Motion    & Wiener Process   \\
    \midrule
    5 & 0.69 , 0.02 & 0.39 , 0.08 & 0.80 , 0.01    \\
    50 & 0.74 , 0.35 & 0.26 , 0.03 & 0.81 , 0.16       \\
    500 & 0.73 , 0.69 & 0.23 , 0.04 & 0.80 , 0.50 \\
    \bottomrule
  \end{tabular}
\end{table}

\begin{table}
  \caption{Performance metrics for time series using retargeting wavelet scalograms}
  \label{retargeting}
  \centering
  \begin{tabular}{ccccccc}
    \toprule
    \multicolumn{3}{r}{Precision , Recall}                   \\
    \cmidrule(r){2-4}
    Size of Training Dataset & Brownian Bridge     &  Drifted Brownian Motion    & Wiener Process   \\
    \midrule
    5 & 0.16 , 0.29 & 0.01 , 0.00 & 0.16 , 0.07   \\
    50 & 0.12 , 0.83 & 0.04 , 0.10 & 0.17 , 0.66      \\
    500 & 0.13 , 0.93 & 0.02 , 0.17 & 0.19 , 0.79 \\
    \bottomrule
  \end{tabular}
\end{table}

\newpage
\section{Conclusion}
A generative framework is presented for timeseries based on wavelet transform and multi-scale patch-based generative architectures. The intuition behind the framework is to capture the recurrence of information within each single instance of time series scalogram to generate similar time series of the same length (reshuffling) and of different length (retargeting). The framework is promising for time series with less trends compared with trended ones and for reshuffling compared with retargeting, according to the fidelity and diveristy performance metrics. More studies and experiments are therefore required to tackle these two issues with trends and retargeting, as well as to benchmark the proposed framework in comparison with other time series generative architectures in low-data regime.

\medskip

\bibliographystyle{unsrt} 
\bibliography{main}

\end{document}